\documentclass[
  preprint,
  amsmath,amssymb,
  aps
]{revtex4-2}

\usepackage[utf8]{inputenc}
\usepackage[english]{babel}

\usepackage{mathtools}
\usepackage{bm}
\usepackage{mathdots}
\usepackage{amsthm}
\usepackage{ytableau}

\usepackage{graphicx}
\usepackage{dcolumn}

\usepackage{braket}
\usepackage{booktabs}
\usepackage{nicefrac}

\DeclareMathOperator{\imm}{Imm}

\newcommand{\dyad}[1]{\Ket{#1}\!\!\Bra{#1}}
\DeclareMathOperator{\tr}{tr}

\newcommand{\GTket}[3]{%
  \Ket{%
    \setlength{\arraycolsep}{1pt}%
    \footnotesize\begin{array}{ccccccc}
      #1 &   & #2\\
         & #3 &
    \end{array}%
  }%
}
\theoremstyle{definition}
\newtheorem{definition}{Definition}
\theoremstyle{plain}
\newtheorem{theorem}{Theorem}

\newtheorem{corollary}{Corollary}

\theoremstyle{remark}
\newtheorem{remark}{Remark}
\newcommand{\ngates}{g}

\newcommand{\matrixus}{\mathsf{U}}

\DeclareMathOperator*{\average}{\mathbb{E}}
\newcommand{\noise}{\mathcal{E}}
\newcommand{\uniop}{\mathcal{U}}
\newcommand{\hilbert}{\mathcal{H}}
\newcommand{\setchannels}{\mathcal{C}}
\newcommand{\nullirrep}{\mathbf{0}}
\newcommand{\ext}{\text{e}}
\newcommand{\param}{p}
\newcommand{\intpartition}{\varkappa}
\newcommand{\koeff}{\kappa}

\usepackage{hyperref}
\usepackage{glossaries}
\setcitestyle{numbers}
\ytableausetup{smalltableaux}

\newacronym{agf}{AGF}{average gate fidelity}
\newacronym{cptp}{CPTP}{completely positive and trace preserving}
\newacronym{cg}{CG}{Clebsch--Gordan}
\newacronym{spam}{SPAM}{state preparation and measurement}
\newacronym{wot}{WOT}{wonderful orthogonality theorem}
\newacronym{rb}{RB}{randomized benchmarking}
\newacronym{gt}{GT}{Gelfand-Tsetlin}

\begin{document}

\title{Kostant relation in filtered randomized benchmarking for passive bosonic devices}

\author{David Amaro-Alcalá}%
\email{david.amaroalcala@savba.sk}
\affiliation{%
Institute of Physics, Slovak Academy of Sciences, D\'ubravsk\'a cesta 9, Bratislava 845~11, Slovakia
}%

\date{\today}
\begin{abstract}
  We aim to reduce the cost of the current bosonic randomized benchmarking proposal. To do this, we introduce two filter functions: one uses immanants, the other uses characters of the special unitary group. These filters avoid computing \gls{cg} coefficients and yield simple variance expressions. The character filter is not only efficient to compute, but also has a constant, low variance. Our filters rely on the same data as the original proposal. We also discuss an example with photon loss and gain. Our numerical evidence shows that a scheme using weak coherent states and intensity measurements can yield estimates close to those obtained without loss or gain. Our work could support simpler platform characterization and streamline data analysis.
\end{abstract}
\maketitle

\section{Introduction}

Characterizing passive bosonic devices is an important step in the development of a continuous-variable quantum computer~\cite{Masada2015,Takeda2017,Yonezu2023,Fukui2022}.
A recent extension of the 
randomized benchmarking
scheme~\cite{Emerson2005,Magesan2011,Knill2008,AmaroAlcal2024,Jafarzadeh2020,helsen2022},
one of the most successful
methods for characterizing finite-dimensional quantum gates, brings this framework to bosonic passive devices~\cite{mirko2025,wilkens2024}.
This scheme estimates a fidelity-like figure of merit of the noise associated with
such devices. It inherits many desirable features of standard randomized
benchmarking. These include robustness to \gls{spam}
errors and a sound theoretical foundation.

However,
the original proposal suffers from two drawbacks.
First,
it requires evaluating matrix permanents,
which are computationally hard to compute~\cite{Valiant1979,Buergisser2000b}.
The required permanents must be determined on a
case-by-case basis because they depend on complex decompositions
involving \gls{cg} coefficients.
Second,
the experimental design is challenging for most laboratories,
as it requires preparing Fock states
and using photon-number-resolving detectors.

In this work,  
we build upon the existing randomized benchmarking framework for  
bosonic devices~\cite{mirko2025} and address both practical and computational limitations.  
We propose an improved protocol that 
keeps the original method's simplicity.
Our method eliminates the need for photon-number-resolving or homodyne  
detection and removes dependence on permanents and \gls{cg}  
coefficients.
The filter expression can also be known in advance,
as it does not depend on the initial state or measurement.
This greatly simplifies the characterization scheme.

Our approach yields a more experimentally accessible characterization scheme that reduces both computational and experimental costs.
We provide a feasibility analysis. Our analysis suggests that experimental data compatible
with the protocol can be obtained using weak coherent states and intensity
measurements. This broadens the practical scope of the method.
We also simplify data analysis by leveraging a relation due to the late Bertram
Kostant~\cite{Kostan1995}.
This relation allows us to express the filter function in terms of a smaller set of less computationally expensive immanants,
avoiding the computation of \gls{cg} coefficients.
In addition, we show that using the characters of the
special unitary group leads to an efficient
filter for any irrep.

This paper is organized as follows.
In Sec.~\ref{sec:background},
we introduce the tools and the problem.
In particular, we recall the relation between immanants,
characters, and irreducible representations (irreps) of the unitary group.
In Sec.~\ref{sec:approach},
we introduce immanant and character filters.
We show that they yield a single exponential decay of the parameters in the figure of
merit.
We also present an alternative to data-gathering process that is
simpler and needs fewer complex experimental arrangements. In
Sec.~\ref{sec:results}, we describe the filtering process in detail,
including loss and gain errors and a numerical test.
Finally,
in the discussion and conclusion section (Sec.~\ref{sec:discussion}), we summarize our
scheme, compare it to the original, and highlight its advantages.

\section{Background}
\label{sec:background}

\subsection{States and channels}
In this subsection,
our goal is to introduce the representation of unitary operators as interferometers.
We start by introducing the Hilbert space and the states.
We then proceed to discuss the action of unitary gates on the states,
which allows us to recall the relevant unitary representations.

Throughout this work,
we consider the Hilbert space of \(n\) 
indistinguishable
photons with \(m\) modes,
which we denote by \(\hilbert^{n}_m\).
The states living in \(\hilbert^{n}_m\) 
are labeled by the occupation numbers \(\mathbf{n} = (n_0, \ldots,
n_{m-1})\);
\(n_i\) denotes how many photons are
in the \(i\)-th mode.
The corresponding state is written as
\begin{equation}\label{eq:occupation_number_state}
\ket{\mathbf{n}}
\coloneqq\left|n_1, \ldots, n_m\right\rangle
\coloneqq 
\left(\prod_{k=1}^m
(n_{k}!)^{-1/2}
(a_k^{\dagger })^{n_k} \right)\ket{\mathbf{0}}       
,
\end{equation}
where \(a_k^{\dagger }\) is the creation operator in the  \(k\)-th mode.

We denote by SU(\(m\)) the group of \(m\times m\) unitary matrices
with unit determinant.
The action of a unitary matrix \(U\)
on a creation operator is
\begin{equation}\label{eq:action-unitary}
\uniop(a^{\dagger}_i) \coloneqq 
\sum_j U_{ji} a^{\dagger}_j.
\end{equation}
Therefore,
the action on state
\(\ket{\mathbf{n}}\)
is implicitly defined as
\begin{equation}\label{eq:action-on-state}
\uniop(\ket{\mathbf{n}})  
\coloneqq 
\left(\prod_{k=1}^m
(n_{k}!)^{-1/2}
\left(\uniop(a_k)^{\dagger }\right)^{n_k} \right)\ket{\mathbf{0}}       
\end{equation}
with the product running over all the input ports.
Building on the action on \(\ket{\mathbf{n}}\),
we define the action of a unitary on states~\(\ket{\mathbf{n}}\bra{\mathbf{m}}\)
as
\begin{equation}
\uniop(\ket{\mathbf{n}})\otimes \uniop(\bra{\mathbf{m}}),
\end{equation}
which we denote by
\begin{equation}\label{eq:representation-gamma}
\Gamma\colon
U\mapsto 
\uniop\otimes\overline{\uniop},
\end{equation}
with \(\overline{\uniop}\)
denoting the complex conjugate of~\(\uniop\).

The main ingredient of the characterization is the gates.
For an arbitrary linear map on \(\hilbert^{n}_m\),
we consider the set of \gls{cptp} mappings,
denoted by \(\setchannels(\hilbert^{n}_m)\),
and we denote its elements by \(\noise\).
In this article, we use the terms ``noise" and ``operator" interchangeably to refer to a channel.
Thus, we use \(\noise\) to represent both.
For clarity and simplicity,
we assume a gate- and time-independent,
Markovian
noise model of the form
\begin{equation}\label{eq:noise-def}
  \widetilde{\uniop} = \noise \circ \uniop
\end{equation}
and~\(\widetilde{\varrho} =
\noise(\varrho)\).
The tilde notation indicates objects affected by noise, 
thus \(\widetilde{\varrho}\) denotes \(\noise'(\varrho)\)
for some channel \(\noise'\) not necessarily equal to \(\noise\).

Lastly, in this subsection,
we recall that the vectorization of a state \(\varrho\)
and measurement  \(E\)
is represented by the vectors \(\ket{\varrho}\)
 and \(\ket{E}\) that satisfy
\begin{equation}\label{eq:proba}
  \operatorname{tr}[E \noise(\varrho)]
  =
  \bra{E} \Gamma(\noise) \ket{\varrho}.
\end{equation}
Both sides of Eq.~\eqref{eq:proba} 
  correspond to the probability of measuring \(E\) after the channel \(\noise\) is applied to \(\varrho\).
In finite-dimensional systems~\cite{Lin2019},
the vectorization operation compatible with the transformation \(\Gamma\) is the
tensor product \(\operatorname{vec}(\varrho)\otimes
\operatorname{vec}(\overline{\varrho})\),
where \(\operatorname{vec}\) corresponds to stacking
the rows 
of the matrix representation of \(\varrho\)
into a single vector.
\subsection{Representation theory}
Irreps of the special unitary group play a fundamental
role in both bosonic randomized benchmarking~\cite{mirko2025} and in this work.
In this subsection,
we introduce standard language in representation theory,
describe the decomposition into irreps of the representation in
Eq.~\eqref{eq:representation-gamma},
and introduce the fidelity-like figure of merit used for characterization.

We start with the decomposition of 
Eq.~\eqref{eq:representation-gamma} into irreducible representations.
The irreps of the special unitary group are labeled by integer partitions \(\lambda =
(\lambda_0, \ldots, \lambda_{m-1})\)
of the number of photons \(n\).
For a more extensive discussion and with similar notation, refer to the appendices in
Ref.~\cite{mirko2025}.
The action of \(\uniop\) is discussed in Eq.~\eqref{eq:action-on-state};
for states with~\(m\) photons
it corresponds to the action of the irreducible representation (irrep) of the
special unitary group SU(\(m\)), labeled
by the partition
\begin{equation}\label{eq:irrep-lambda-sym}
\lambda \coloneqq  (n, \underbrace{0,\ldots, 0}_{m-1 \text{ times}}),
\end{equation}
with \(\lambda^{*}\) the dual irrep of \(\lambda\);
an example is provided in Appendix~\ref{app:decomposition}.
Thus, using Pieri's formula~\cite{Fulton1996},
the representation \(\Gamma\) introduced in Eq.~\eqref{eq:representation-gamma}
is reducible and decomposes into a finite list of representations (with no
repetition):
\begin{equation}\label{eq:irrep-decomposition-gamma}
\Gamma \coloneqq  \lambda \otimes \lambda^{*} \cong
\bigoplus_\mu \mu.
\end{equation}
The sum is over the partitions \(\mu\),
calculated by adding~\(n\) elements to the partition \(\lambda^{*}\) in
different columns.
We discuss this in detail in Appendix~\ref{app:decomposition},
where we also discuss the computation of \(\lambda^{*}\) from \(\lambda\).

We note that,
because the physical system corresponds to indistinguishable photons,
the Hilbert space \(\hilbert^{n}_m\)
is isomorphic to the (symmetric) irrep \(\lambda\).
This relation is denoted as
\begin{equation}\label{eq:iso}
\hilbert^{n}_m \cong \lambda
.
\end{equation}
The equivalence in Eq.~\eqref{eq:iso}
is made explicit in Eq.~\eqref{eq:equivalence}.
We recall that a symmetric irrep corresponds to the irrep labeled by a
horizontal (single row) Young tableau.
Therefore,
\(\Gamma\) in Eq.~\eqref{eq:representation-gamma} is isomorphic to the set of linear operators acting on \(\hilbert^{n}_m\).
In other words,
talking about~\(\lambda\) is equal to talking about \(\hilbert^{n}_m\);
likewise,
each channel acting on \(\hilbert^{n}_m\)
is an element of ~\(\Gamma\).

We now recall the figure of merit to characterize the noise \(\noise\)
proposed for the original scheme, which we also use~\cite{mirko2025}.
We emphasize that the noise appears under the assumption that
the noisy gates are of the form specified in Eq.~\eqref{eq:noise-def}.
The goal of the scheme is to estimate the following figure of merit
based on the trace over each symmetry subspace~\(\mu\):
\begin{equation}\label{eq:average-gate-fidelity}
  F(\mathcal{E}) \coloneqq
d_\lambda ^{-2}
\sum_{\mu\in\Gamma} d_\mu \param_\mu(\noise),
\end{equation}
where
\(\Gamma\) is introduced in Eq.~\eqref{eq:representation-gamma},
\( d_{\lambda} \) denotes the dimension of~\(\lambda\),
\(\mu\) is defined in Eq.~\eqref{eq:irrep-decomposition-gamma},
and~\(\param_\mu\) is the trace (divided by the dimension of \(\mu\)) of
\(\Gamma\)
restricted to a basis for the irrep \(\mu\).

\subsection{Bosonic randomized benchmarking}
\label{sec:bosonic-rb}

In this subsection,
we revisit the bosonic \gls{rb} scheme, upon which our own work builds.
In Appendix~\ref{app:decomposition},
we explain how the filtering process works;
that is, we show how to calculate the parameters of the figure of merit in
Eq.~\eqref{eq:average-gate-fidelity} of the noise \(\noise\).
Lastly,
we also define the quantities our scheme requires.

Consider a system of \(n\) photons 
accessing
(simultaneously~\cite{AmaroAlcal2020}) an \(m\)-mode interferometer.
Likewise, consider the initial state \(\varrho\) and the measurement \(E\).
The state \(\varrho\) undergoes a transformation by one of \(K\) 
sequences (uniformly randomly sampled) of gates
\begin{equation}\label{eq:notation-sequence gates}
\mathbf{U}^{\ngates}_s
\coloneqq 
(U_0(s), \ldots, U_{\ngates-1}(s)),
\end{equation}
with \(\ngates\) the depth (number of gates) of the sequence,
and~\(s\in \{1,\ldots, K\}\) is the index of the sequence;
for convenience, we drop the dependence on \(s\).
For concreteness and to avoid introducing more notation,
we consider \(g\) taking values from~\(1\) to \(L\),
with \(L\) the maximum circuit depth.

We recall the definition of the filter for the bosonic \gls{rb} scheme.
Then the original filter requires the computation of the following quantity:
\begin{equation}\label{eq:eq:original-filter}
f_{\lambda, \text{orig}}
\coloneqq 
\bra{\varrho}
P_\lambda S^{+} \Gamma(\mathbf{U}^{\ngates}_s)
\ket{E}.
\end{equation}
We explain the quantities that appear as follows:
\(S^{+}\) is the Penrose pseudo-inverse of  
\begin{equation}
  S \coloneqq  \average_U\left[ \Gamma(U)^{\dagger}
\ket{E}\!\!\bra{\widetilde{E}}
\Gamma(U)\right],
\end{equation}
where \(\average_U\) denotes the average over SU(\(m\)),
\(\widetilde{E}\) represents the noisy version (an unknown \gls{cptp}
mapping acting on it) of the measurement \(E\),
and \(P_\mu\) is the projector onto the irrep in the decomposition of~\(\mu\in\Gamma\).
In subsequent applications, we omit the square brackets after using \(\average_U\).
In Section~\ref{sec:approach},
we provide a detailed description of the procedure
for comparison with our filter.

We conclude this subsection with a brief commentary on a recurring assertion across various works.
It has been stated that \gls{rb} encounters the so-called gauge-freedom
issue~\cite{proctor2017}.
The first thing to note is that this happens only for coherent noise.
It is, however,
known that employing randomized compiling (RC)
mitigates this situation~\cite{wallman2016,Hashim2021}.
The reason is that carrying out RC and \gls{rb} together,
coherent noise is mapped into incoherent noise.
Thus, RC effectively addresses the so-called gauge issue.

\subsection{Kostant relation: immanants and zero-weight states}
\label{sub:kostant-immanant-zero}

This manuscript’s core contribution requires discussing immanants and D-functions.
The discussion begins by introducing \gls{gt} patterns,
which help assign unique quantum numbers to the states in \(\lambda\).
The next section connects these patterns to the Fock basis.
Several quantities required for Kostant’s relation are then calculated.
Subsequently,
the text reviews the definition of immanants,
including the characters of the symmetric group.
The manuscript concludes with an explanation of Kostant’s relation.

Whereas the states of the Hilbert space
\(\hilbert^{n}_m\)
lie
in the symmetric irrep
\(\lambda\) of the
unitary group, 
we make use of other irreps appearing in the tensor product 
of \(\lambda\) and~\(\lambda^{*}\); that is,
the dual of the irrep \(\lambda\).
We use \(\lambda\) to denote 
the irrep associated with the unitary evolution in Eq.~\eqref{eq:action-on-state}
and~\(\mu\) to denote the irreps that appear in the decomposition of~\(\lambda \otimes \lambda^{*}\).
This clarification is important because \(\lambda\) is used as part of state and measurement labels,
whereas~\(\mu\) is part of the label of zero-weight states necessary for describing our filter.

Input and output states of a configuration with \(n\) photons entering
simultaneously into
an \(m\)-port interferometer are labeled by \gls{gt} patterns.
The reason is that these patterns label the states for irreducible
representations of the unitary group~\cite{raczka1986}.
Therefore, due to the isomorphism mentioned in Eq.~\eqref{eq:iso},
we can also label the system's states using these patterns.
Each pattern is an array with \(m\) rows,
where each row has decreasing length:
\begin{equation}\label{eq:gt-pattern}
M \coloneqq 
{\footnotesize
\setlength{\arraycolsep}{1pt}
\begin{array}{ccccccc}
M_{1,1} &         & M_{1,2} &         & \cdots  &           & M_{1,m} \\
        & M_{2,1} &         & \cdots  &         & M_{2,m-1} & \\
        &         & \ddots  &         & \iddots &           & \\
        &         &         & M_{m,m} &         &           &
\end{array}}.
\end{equation}
The first row is equal to \(\lambda\),
introduced in Eq.~\eqref{eq:irrep-lambda-sym},
the integer partition labeling the
irrep,
and the remaining rows can be computed using the so-called betweenness condition:
\begin{equation}
M_{i,j} \geq M_{i+1,j} \geq M_{i,j+1}\geq 0
.\end{equation} 

We now discuss  how to represent the states from~\(\hilbert^{n}_m\) using
\gls{gt} patterns.
For the Fock state 
\(\ket{\mathbf{n}} = \ket{n_1,\ldots, n_m}\),
the following \gls{gt}~pattern is
assigned:
\begin{equation}\label{eq:assign-state-pattern}
N  = 
\footnotesize
\setlength{\arraycolsep}{1pt}
\begin{array}{ccccccc}
\sum_{i=1}^{m} n_i &                      & 0      &        & \cdots  &   & 0       \\
                   & \sum_{i=1}^{m-1} n_i &        & \cdots &         & 0 & \\
                   &                      & \ddots &        & \iddots &   & \\
                   &                      &        & n_1    &         &   &
\end{array}.
\end{equation}
Thus, 
we can write the following assignment
\begin{equation}\label{eq:equivalence}
\ket{\mathbf{n}}\cong\ket{N},
\end{equation}
where \(\ket{N}\) uses \(N\) of Eq.~\eqref{eq:assign-state-pattern}.
Notice that the equivalence of Eq.~\eqref{eq:equivalence}
stems from Eq.~\eqref{eq:iso}.

The next topic to discuss before describing Kostant’s result is zero-weight
states for irreps of the unitary group.
Consider an irrep \(\mu\) of SU(\(m\)).
We first need the occupation number of a state as
\begin{equation}\label{eq:occ-number}
n_i(M) \coloneqq  \sum_j M_{i,j} - \sum_{j'} M_{i+1, j'}.
\end{equation}
The weight of a state is, in turn, defined in
terms of the occupation number.
Then, the weight is defined
as the difference between adjoining occupation numbers: 
\begin{equation}\label{eq:weight}
\mathbf{w}_N = (n_1-n_2, n_2-n_3, \ldots, n_{m}- m_{m-1}).
\end{equation}
\begin{definition}(Zero-weight states in the basis of an irrep)\label{def:zero-weight}
Consider an irrep \(\mu\) of SU(\(m\)).
Let \(\ket{M}\) be a basis state of \(\mu\) labeled
with the \gls{gt} pattern of Eq.~\eqref{eq:gt-pattern}.
Then,~\(\ket{M}\) is a zero-weight state 
if
every entry of \(\mathbf{w}_M\) is equal to zero.
The basis elements of \(\mu\) with zero weight are denoted by \(\mathcal{Z}_\mu\).
\end{definition}
\noindent Two examples of zero-weight states are described in
Appendix~\ref{app:kostant-relation-example},
Eqs.~\eqref{eq:subeq-zero-weight-states-example}.

Immanants,
the second ingredient needed to introduce Kostant's relation,
are generalizations of the determinant and the permanent~\cite{littlewood1977}.
These numerical quantities are maps from the set of matrices to a complex
number.
Let~$\intpartition$ be an integer partition of $m$,
and~$\chi_{\intpartition}(\sigma)$ denote the character for the group element $\sigma$ of
the symmetric group for \(m\) elements \(\mathfrak{S}_m\)~\cite{littlewood1977}.
Then the immanant is defined as
\begin{equation}\label{eq:immanant}
\imm_\intpartition(U) \coloneqq 
\sum_{\sigma\in\mathfrak{S}_m} \chi_{\intpartition}(\sigma) \prod_{i=1}^{m} U_{i,\sigma(i)},
\end{equation}
where $\sigma(i)$ denotes a permutation of the value of $i$ according to $\sigma$.

We illustrate the immanant using well-known cases,
then present the first uncommon one.
First, we provide a character table for \(\mathfrak{S}_2\)
and \(\mathfrak{S}_3\).
\begin{table}[h!]
\centering
\caption{\label{tab:char-table}Character tables of $\mathfrak{S}_2$ and $\mathfrak{S}_3$.}
\vspace{1em}

\begin{tabular}{c|cc}
\multicolumn{3}{c}{\textbf{$\mathfrak{S}_2$}} \\
\toprule
\(\mu\)  & $e$ & $(12)$ \\
\midrule
$(2,0)$  & 1   & 1 \\
$(1,1)$  & 1   & $-1$ \\
\bottomrule
\end{tabular}

\vspace{1.5em}

\begin{tabular}{c|cccccc}
\multicolumn{7}{c}{\textbf{$\mathfrak{S}_3$}} \\
\toprule
\(\mu\)   & $e$ & $(12)$ & $(13)$ & $(23)$ & $(123)$ & $(132)$ \\
\midrule
$(3,0,0)$ & 1   & 1      & 1      & 1      & 1       & 1 \\
$(2,1,0)$ & 2   & 0      & 0      & 0      & $-1$    & $-1$ \\
$(1,1,1)$ & 1   & $-1$   & $-1$   & $-1$   & 1       & 1 \\
\bottomrule
\end{tabular}

\end{table}
Then, using
Eq.~\eqref{eq:immanant} and Table~\ref{tab:char-table},
for \(\mathfrak{S}_2\), we have two immanants:
\begin{subequations}
  \begin{align}
    \imm_{(2,0)}(U) &= U_{11}U_{22} + U_{12}U_{21} = \operatorname{per}(U),\\
    \imm_{(1,1)}(U) &=U_{11}U_{22} - U_{12}U_{21} = \det(U).
  \end{align}
\end{subequations}

Thus, these two quantities are already known.
Next, for \(\mathfrak{S}_3\) using the
Table~\ref{tab:char-table}, we have the following immanants for a matrix
\(U\):
\begin{subequations}
  \begin{align}
    \imm_{(3,0,0)}(U)
&= U_{11}U_{22}U_{33}
+ U_{12}U_{23}U_{31} \nonumber\\
&\quad+ U_{13}U_{21}U_{32}
+ U_{12}U_{21}U_{33}\nonumber\\
&\quad+ U_{13}U_{22}U_{31}
+ U_{11}U_{23}U_{32},\\
\imm_{(1,1,1)}(U)
&= U_{11}U_{22}U_{33}
+ U_{12}U_{23}U_{31} \nonumber\\
&\quad+ U_{13}U_{21}U_{32}
- U_{12}U_{21}U_{33}\nonumber\\
&\quad- U_{13}U_{22}U_{31}
- U_{11}U_{23}U_{32},\\
\imm_{(2,1,0)}(U)
&= 2\,U_{11}U_{22}U_{33}
- U_{12}U_{23}U_{31}\nonumber\\
&\quad- U_{13}U_{21}U_{32}\label{eq:imm-novel}.
  \end{align}
\end{subequations}

The immanant of Eq.~\eqref{eq:imm-novel}
is the first immanant which is
neither a permanent nor a determinant.
We conclude this comment on immanants by noting that
there is now a Wolfram package~\cite{WolframImmanant} that
computes these quantities,
without having to look up a character table.
In Theorem~\ref{thm:filter},
we use these non-determinant and non-permanent immanants in
our filter procedure, where we show numerically how to perform 
the data~analysis.

The relation by Kostant refers to 
the fact that an immanant can be computed from the states
with zero-weight of a given irrep~\cite{Kostan1995,Guise2016}. 
Computing the trace over the state with weight zero of a
representation containing a single copy of the irrep is equal to the immanant~$\lambda$ of the fundamental representation.
  \begin{theorem}[Kostant relation~\cite{Kostan1995,deguise2018}]\label{thm:kostant}
Let \(\imm_\intpartition(U)\)
denote the immanant of \(U\) corresponding to the partition~\(\intpartition\),
which is introduced in Eq.~\eqref{eq:immanant}.
Then,
\begin{equation}
\sum_{\ket{\zeta_\intpartition}\in \mathcal{Z}_\intpartition}
\bra{\zeta_\intpartition} \Gamma(U) \ket{\zeta_\intpartition}
=
\imm_\intpartition(U),
\end{equation}
with the states \(\ket{\zeta_\intpartition}\) introduced in Definition~\ref{def:zero-weight}.
\end{theorem}
In Appendix~\ref{app:kostant-relation-example},
we show an example for SU(3) comparing the immanant with the sum over
zero-weight states D-functions.
In Sec.~\ref{sec:results},
we show how Theorem~\ref{thm:kostant} is used
to eliminate the need to compute \gls{cg} coefficients
and multiple matrix permanents.

\subsection{Characters of SU(\(m\))}\label{sub:su-characters}

\begin{definition}[Character of SU(\(m\))]\label{def:su-character}
Let \(\mu\) be an irrep of SU(\(m\)).
The character \(\chi_\mu\) is the trace of the representation \(\Gamma\) over a
basis of \(\mu\), namely
\begin{equation}\label{eq:su-character}
  \chi_\mu(U)\coloneqq
  \sum_{\ket{\psi} \in \mathcal{B}_\mu}
  \bra{\psi} \Gamma(U) \ket{\psi},
\end{equation}
with \(\mathcal{B}_\mu\) an orthonormal basis for \(\mu\).
\end{definition}
\noindent To avoid ambiguity, hereinafter we use the term character only for those of
SU(\(m\)).
In Appendix~\ref{app:characters}, we recall several examples of character
computations using this definition. Although the Weyl determinant formula seems
simple~\cite{Fulton1996}, evaluating it with floating-point arithmetic can be
problematic because it involves determinants of matrices with entries of
similar magnitude. Therefore, it is preferable to use the Schur-polynomial
form.

With the notation already introduced, we can define the parameters \(p_\mu\)
of Eq.~\eqref{eq:average-gate-fidelity} more explicitly:
\begin{equation}\label{eq:parameters}
p_\mu(\noise)
\coloneqq
  \sum_{\ket{\psi} \in \mathcal{B}_\mu}
  \bra{\psi} \Gamma(\noise) \ket{\psi}.
\end{equation}

To summarize Sec.~\ref{sec:background},
we have recalled important notions for passive photonic devices.
We also recalled relevant SU(\(m\)) representation-theoretic
structure. We reviewed immanants, Kostant’s relation, and SU(\(m\))
characters.
We now use these concepts in a filtered benchmarking scheme.

\section{Approach}\label{sec:approach}

This section presents a concise way to describe the sequence of gates needed for benchmarking.
We then use this method to recall the filter definition from the original proposal.  
Furthermore,
we present our own filters and demonstrate that employing them results in a single exponential for a parameter.
It is important to note that our scheme can use the same data as the original filter.

Our first task is to define what filtering means.
Doing so clarifies not only
the original scheme but also our contribution.
We begin by introducing notation for a sequence of gates.
Consider an ordered sequence of
\(g\)
gates~\(\mathbf{U}^{\ngates}_s\),
an initial state \(\varrho\), and a measurement \(E\).
Further,
consider that each \(U\) 
in \(\mathbf{U}^{\ngates}_s\) is an element of the unitary group acting on
\(\hilbert^{n}_m\).
Similarly,
denote the \(s\)-th randomly sampled sequence of~\(g\) gates by \(\mathbf{U}^{\ngates}_s\).
We collect all these sequences into the following matrix:
\begin{equation}
\matrixus \coloneqq 
\begin{bmatrix}
\mathbf{U}^{\ngates=1}_{s=1} & \mathbf{U}^{\ngates=2}_{s=1} & \dots  & \mathbf{U}^{\ngates=K}_{s=1} & \\
\mathbf{U}^{\ngates=1}_{s=2} & \mathbf{U}^{\ngates=2}_{s=2} & \dots  & \mathbf{U}^{\ngates=K}_{s=2} & \\
\vdots                       & \vdots                       & \ddots & \vdots                       & \\
\mathbf{U}^{\ngates=1}_{s=N} & \mathbf{U}^{\ngates=2}_{s=N} & \dots  & \mathbf{U}^{\ngates=K}_{s=N} & \\
\end{bmatrix},
\end{equation}
where the entry \(\matrixus_{\ngates,s}\) represents the sequence of gates~\(\mathbf{U}^{\ngates}_{s}\).
We incur, for convenience,
in the following abuse of notation:
\(\Gamma(\mathbf{U}^{\ngates}_{s})
\coloneqq 
\Gamma(\prod_g \mathbf{U}^{\ngates}_{s})
\).
Likewise, the 
real-world experimental data 
is written as
\begin{multline}\label{eq:experimental_data}
  d^{(\ngates,s)}(\widetilde{\matrixus}_{\ngates,s})
\coloneqq 
\bra{\widetilde{E}} \Gamma(\widetilde{\matrixus}_{\ngates,s}) \ket{\widetilde{\varrho}}
\\=
\operatorname{tr}[\widetilde{E}
\bigcirc_{U\in \matrixus_{\ngates,s}}\widetilde{\mathcal{U}} (\widetilde{\varrho}) ],
\end{multline}
where \(\bigcirc\) denotes composition.
We group
\(d^{(\ngates,s)}(\widetilde{\matrixus}_{\ngates,s})\)
into a matrix:
\begin{equation}
  \label{eq:data-matrix}
\mathsf{D} \coloneqq 
\begin{bmatrix}
d^{(1,1)} & d^{(2,1)} & \dots  & d^{(K,1)} & \\
d^{(1,2)} & d^{(2,2)} & \dots  & d^{(K,2)} & \\
\vdots    & \vdots    & \ddots & \vdots    & \\
d^{(1,L)} & d^{(2,L)} & \dots  & d^{(K,L)} & \\
\end{bmatrix}.
\end{equation}
Now we proceed with our description of the filtering process.

Using a filtering process,
we estimate every parameter~\(\param_\mu\) to calculate \(F(\noise)\)
in Eq.~\eqref{eq:average-gate-fidelity},
which is a function only of the parameters \(\param_\mu\) and \(d_\lambda\).
Our proposed filter function is described in the following theorem.
\begin{theorem}[Immanant filter function]
  \label{thm:filter}
  Let
\begin{equation}\label{eq:new-filter}
f_{\imm, \mu}^{(g,s)}(\mathsf{U}_{g,s})
\coloneqq 
\imm_\mu(\mathsf{U}_{g,s})
\end{equation}
be our filter function.
Then,
\begin{equation}\label{eq:filtering-process}
  \Phi_g^{(f)} \coloneqq 
  \average_{s}
  f^{(\ngates,s)}_{\mu}(\matrixus_{\ngates,s})
  d^{(\ngates,s)}_{\varrho, E}(\widetilde{\matrixus}_{\ngates,s})
  = 
  \koeff p_\mu^{g-1},
\end{equation}
\end{theorem}
\noindent for some constant \(\koeff\), 
which is irrelevant to the characterization.
Thus,
our filter function isolates a single parameter
and can be used to estimate \(F(\noise)\) of Eq.~\eqref{eq:average-gate-fidelity}.
\begin{proof}
We now explicitly justify the form of our filter.
The summary of the proof consists of first summing over zero-weight states and then averaging over ordered sequences of \(g+1\) gates; \(g\) gates are used for the twirling of the noise and another for an auxiliary twirling related to the noisy measurement.
We begin with the sum over zero-weight states:
\begin{equation}
\sum_{\ket{\zeta_\mu} \in \mathcal{Z}_\mu}
\bra{\zeta_\mu} \Gamma(\mathsf{U}_{g,s})^{\dagger} \ket{\zeta_\mu} 
\bra{\widetilde{E}} 
\Gamma(\widetilde{\mathsf{U}}_{g,s})\ket{\widetilde{\varrho}}.
\end{equation}
Consider \(\ket{\zeta^{(i)}_\mu}\) the \(i\)-th zero-weight state in \(\mathcal{Z}_\mu\);
these are orthogonal vectors.
Likewise,
consider the twirled operator 
\begin{equation}\label{eq:definition-operator-s}
S_{\imm_\mu}^{(i)}
\coloneqq 
\average_{U\in \mathrm{SU}(m)}
\Gamma(U)^{\dagger} \ket{\zeta_\mu^{(i)}} 
\bra{\widetilde{E}} 
\Gamma(U)
\end{equation}
(for each zero-weight state);
notice that \(S_{\imm_\mu}^{(i)}\) is defined for a single gate, not for a sequence.
Now,
since 
\(\mathbf{U}^{g+1}_s\)
is a sequence of \(\ngates+1\) gates:
\begin{multline}\label{eq:final-expression}
\average_{\mathbf{U}^{\ngates+1}}
\sum_{\ket{\zeta_\mu^{(i)}} \in \mathcal{Z}_\mu}
\bra{\zeta_\mu^{(i)}} 
\Gamma(\mathsf{U}_{g+1,s})^{\dagger}
\ket{\zeta_\mu^{(i)}} 
\bra{\widetilde{E}} 
\Gamma(\widetilde{\mathsf{U}}_{g+1,s})
\ket{\widetilde{\varrho}}
\\=
\sum_{\ket{\zeta_\mu^{(i)}}}
\bra{\zeta_\mu^{(i)}} 
S_{\imm_\mu}^{(i)} T[\noise]^{\ngates}
\ket{\widetilde{\varrho}} 
,
\end{multline}
where
\(\average_{\mathbf{U}^{\ngates+1}}\)
denotes the uniform average over
every  multiset with length \(\ngates+1\)
and
\begin{equation}\label{eq:twirl-definition}
  T[\noise] \coloneqq 
  \average_U \Gamma(U)^{\dagger} \Gamma(\noise) \Gamma(U).
\end{equation}
As we now discuss,
Eq.~\eqref{eq:final-expression}
reveals that we could have a single exponential.
First,
note that both $S_{\imm_\lambda}^{(i)}$ and $T[\noise]$ have the same irrep decomposition.

Since each~$\ket{\zeta_\mu^{(i)}}\in \mu$ of Eq.~\eqref{eq:irrep-decomposition-gamma},
each term~$\bra{\zeta_\mu} S_{\imm_\lambda}^{(i)} T[\noise]\ket{\widetilde{\varrho}}$
is proportional to~$\param_\mu$,
introduced in Eq.~\eqref{eq:average-gate-fidelity}.
We demonstrate that as follows.
Notice that 
\(S_{\imm_\mu}^{(i)}\)
does not mix isotypic components:
\begin{equation}\label{eq:coefficients-s}
  \bra{\zeta_\mu^{(i)}} S_{\imm_\mu}^{(i)} =
  \sum_j s_{i,j} \bra{\zeta_\mu^{(j)}},
\end{equation}
then, because \(T[\noise]\) is a direct
sum of homotheties (multiples of the identity map),
\begin{equation}
  \bra{\zeta_\mu^{(i)}} S_{\imm_\mu}^{(i)}T[\noise]^{g}=
  \left(\sum_j s_{i,j}\bra{\zeta_\mu^{(j)}}\right) \param_\mu^{g}.
\end{equation}
To conclude, we note that
\begin{equation}
  \bra{\zeta_\mu^{(i)}} S_{\imm_\mu}^{(i)}T[\noise]^{g} \ket{\widetilde{\varrho}}=
  \left(\sum_j s_{i,j}\langle\zeta_\mu^{(j)} | \widetilde{\varrho}\rangle\right) \param_\mu^{g}.
\end{equation}
Thus, setting the constant \(\koeff \coloneqq \left(\sum_j
s_{i,j}\langle\zeta_\mu^{(j)} | \widetilde{\varrho}\rangle\right)\),
we conclude the proof.
\end{proof}
This shows that, despite the sum,
we still obtain a single exponential decay
for a single parameter \(\param_\mu\).
Notice that this proof is valid for any D-function.
Immanants are obtained by summing the diagonal D-functions over zero-weight
states.  Extending the sum over all the diagonal D-functions, we calculate the
character for such a group element in that irrep.
Thus, since we again are dealing with sums of D-functions
restricted to some irrep, the observations used in the proof of Theorem~\ref{thm:filter} also
apply to characters.

\begin{corollary}[Character filter]\label{cor:character-filter}
Let
\begin{equation}\label{eq:char-filter}
f_{\chi,\mu}^{(g,s)}(\mathsf{U}_{g,s})
\coloneqq
\chi_\mu(\mathsf{U}_{g,s})
\end{equation}
be the filter function defined by the character of an irrep~\(\mu\) of SU(\(m\))
introduced in Eq.~\eqref{eq:su-character}.
Then,
\begin{equation}\label{eq:char-filtering-process}
  \Phi_g^{(\chi)} \coloneqq 
  \average_{s}
  f^{(\ngates,s)}_{\chi,\mu}(\matrixus_{\ngates,s})
  d^{(\ngates,s)}_{\varrho, E}(\widetilde{\matrixus}_{\ngates,s})
  = 
  \koeff_\chi p_\mu^{g-1},
\end{equation}
for some constant \(\koeff_\chi\).
\end{corollary}
We now recall two facts relevant to the assessment of the feasibility of the scheme in terms of sample complexity
and computational complexity.
\begin{remark}[Character orthogonality and variance]\label{rem:char-variance}
  From the \gls{wot} we obtain
  \begin{equation}
    \int_{\mathrm{SU}(m)}\chi_\mu(U)\overline{\chi}_\nu(U)\,\mathrm{d}U=\delta_{\mu\nu} ,
  \end{equation}
with \(\mathrm{d}U\) denoting the Haar measure.
This immediately gives
$\mathbb{E}[\chi_\mu(U)]=0$ for nontrivial \(\mu\),
where the average is computed by randomly sampling from the Haar measure a
unitary matrix \(U\) and evaluating the character.
Using again the \gls{wot} we get
$\mathrm{Var}(\chi_\mu(U))=\mathbb{E}[|\chi_\mu(U)|^2]=1$, i.e., constant
variance independent of the irrep and group.
\end{remark}
\begin{remark}[Proposition~7.4 in~\cite{Buergisser2000a}]
  \label{rem:efficient}
The evaluation of~SU(\(m\)) characters is efficient for any irrep.
Numerical character evaluation is implemented in
\cite{GroupFunctionsjl}.
In particular,  the
character 
\(\chi_\mu(U)\)
can be evaluated with a number of
arithmetic operations polynomial in \(\max\{\mu_1,m\}\).
\end{remark}

Remarks~\ref{rem:char-variance}
and~\ref{rem:efficient}
show that, among the filters considered here, the character filter has the most
favorable computational and sampling properties. In particular, it combines
polynomial-time evaluation with known constant variance.

	\section{Results}
	\label{sec:results}

In this section, 
we analyze loss and gain errors, considering the feasibility of applying our filter in scenarios where the Hilbert space is not limited to a fixed photon number.
We focus on weak coherent states, which can be prepared more readily than other
states, and on intensity measurements.
Finally, we compare the computational cost of our filter with that of the original.

\subsection{Filtering process}

Assume the experiment involves \(n\) photons and the interferometer has \(m\) ports.
With Theorem~\ref{thm:filter} and Corollary~\ref{cor:character-filter} at hand,
we can now explain how our filter estimates the fidelity-like quantity.
Let \(f\) denote either the immanant filter \(f_{\imm,\mu}\) of
Eq.~\eqref{eq:new-filter} or the character filter~\(f_{\chi,\mu}\) of
Eq.~\eqref{eq:char-filter}; the latter is computationally more efficient, as
discussed after Corollary~\ref{cor:character-filter} and in
Remarks~\ref{rem:char-variance} and~\ref{rem:efficient}.
We gather data in the matrix
\(\mathsf{D}\) of Eq.~\eqref{eq:data-matrix}.
We label it based on the sequence number \(s\) and the circuit depth \(g\) used.
It is crucial to keep track of the sequence~\(\mathsf{U}_{g,s}\) used.
Then, the corresponding filter values are computed for each sequence, and we organize them into a matrix
\begin{multline}\label{eq:matrix-filter}
  \mathsf{F}^{(f)}_\mu
=\\
\begin{bmatrix}
f^{(1,1)}_\mu(\mathsf{U}_{1,1}) & f^{(2,1)}_\mu(\mathsf{U}_{2,1}) & \dots  & f^{(K,1)}_\mu(\mathsf{U}_{K,1}) & \\
f^{(1,2)}_\mu(\mathsf{U}_{1,2}) & f^{(2,2)}_\mu(\mathsf{U}_{2,2}) & \dots  & f^{(K,2)}_\mu(\mathsf{U}_{K,2}) & \\
\vdots                          & \vdots                          & \ddots & \vdots                          & \\
f^{(1,L)}_\mu(\mathsf{U}_{1,L}) & f^{(2,L)}_\mu(\mathsf{U}_{2,L}) & \dots  & f^{(K,L)}_\mu(\mathsf{U}_{K,L}) & \\ \end{bmatrix}
,
\end{multline}
where \(K\) denotes the maximum circuit depth to use, and~\(L\) is the number of different
circuits used.
To estimate each parameter \(\param_\mu\), we use the matrices in
Eqs.~\eqref{eq:data-matrix} and \eqref{eq:matrix-filter} to
compute the following Hadamard product\footnote{
\((A\odot B)_{i,j} \coloneqq A_{i,j} B_{i,j}\).
} 
\begin{equation}
\boldsymbol{\Phi}_g^{(f)} = \sum_s (\mathsf{F}^{(f)}_\mu \odot \mathsf{D})_{s,g}.
\end{equation}
Up to the overall normalization by the number of sampled sequences, this is the
same filtered quantity that appears in Theorem~\ref{thm:filter} for the
immanant filter and in Corollary~\ref{cor:character-filter} for the character
filter. Therefore,
\begin{equation}
\boldsymbol{\Phi}_g^{(f)}
\varpropto p_\mu^{g-1}.
\end{equation}
By fitting an exponential to 
the graph
\(\{g, \boldsymbol{\Phi}_g^{(f)}\}\),
we estimate the parameter \(p_\mu\).
Repeating this process for each irrep \(\mu\),
we then use Eq.~\ref{eq:average-gate-fidelity} to compute the fidelity-like quantity
\(F(\noise)\). 
This concludes our presentation of the filtering process.
In the numerical study below, we specialize in the character filter.
We now describe how this scheme can be analyzed, at the level of feasibility,
with weak coherent states and intensity measurements.

\subsection{Gain and loss noise}
We now demonstrate that our filter can be used to estimate \(F(\noise)\) even in
cases where the noise acts on other Hilbert spaces,
corresponding to the gain or loss of photons.
Beyond extending the interest of our scheme,
this also motivates studying whether weak coherent states and intensity
measurements can provide useful data for the characterization.

Now we consider an extended Hilbert space,
corresponding to the direct sum over the spaces with an arbitrary number of
photons:~\((\hilbert_\ext)^{n}_m \coloneqq  \oplus_{n\geq 0} \hilbert^{n}_m\),
with \(\Gamma_\ext\) being the notation for the unitary action on \((\hilbert_\ext)^{n}_m\).
We use the subindex “e” to denote the representation acting on that system.
Likewise, we use two extensions (including other Hilbert spaces using direct sum) for unitary operations.
The first one extends an operator using the identity to the noisy space,
and the other extends with the null operator (maps every vector
to the null vector of the vector space):
the first is denoted as \(\Gamma_{\text{e}, I}\),
and the other as \(\Gamma_{\text{e}, \varnothing}\).
We now demonstrate that the same steps used in the original scheme remain unchanged.

By Schur’s lemma,
for any operator \(\Gamma_\ext(\noise)\),
the extended operator
\begin{equation}
  S_\ext \coloneqq 
  \average_U \Gamma_{\ext, I}^{\dagger}(U) \Gamma(\noise) \Gamma_{\ext,
  \varnothing}(U)
\end{equation}
has support only in \(\hilbert^{n}_m\);
that is, for any \(\ket{\varrho}\in (\hilbert_\ext)^{n}_m\),
we have \(S_\ext \ket{\varrho} \in \hilbert^{n}_m\).
Note that \(\noise\) already acts on the extended Hilbert space.
A corollary of this observation is as follows.
In the setting considered here, weak coherent states and intensity measurements
may provide data suitable for the filtering procedure.
We examine this feasibility claim below.

The first modification is to use a single coherent state as input \(\varrho\),
in particular, a weak coherent state, which is simpler to prepare and more
readily available than a Fock state~\cite{want2019}.
Consider the case of one weak coherent state entering a beam splitter.
Thus,
there are two modes.
The ideal state (up to normalization), in the occupation-number basis, is
\begin{equation}
\ket{\varrho} = \ket{(0,0)} + \alpha \ket{(1,0)} + O(|\alpha|^2).
\end{equation}
We translate the 
label of the state from the occupation number to the \gls{gt}~pattern basis using Eq.~\eqref{eq:assign-state-pattern}:
\begin{equation}
\ket{\varrho} \cong
\GTket{0}{0}{0}
+
\alpha
\GTket{1}{0}{1}
+ O(|\alpha|^2).
\end{equation}
We note that the state belongs to two different irreps:
spin~zero and spin~one.
Therefore,
the set of channels on the extended Hilbert space
decomposes as:
\begin{multline}\label{eq:decom-coherent}
\setchannels(\hilbert_\ext)\cong
  (\nullirrep \oplus \ydiagram{1})\otimes
  (\nullirrep \oplus \ydiagram{1}^{*})
  \\=
\nullirrep
\oplus \ydiagram{1}
\oplus \ydiagram{1}
\oplus \ydiagram{2}
\oplus \ydiagram{1,1},
\end{multline}
where \({}^{*}\) denotes the dual irrep (see Appendix~\ref{app:decomposition})
and~\(\nullirrep\) the spin zero irrep.
Thus, \emph{a priori}, it is unclear whether the filtering process applies.
We argue that this is indeed the case.

From the decomposition of Eq.~\eqref{eq:decom-coherent},
we see that the irrep corresponding to the parameter \(p_\mu = p_{\ydiagram{2}}\),
which the original scheme aims to estimate,
appears once.
Thus,
by the orthogonality of irreps,
the character filter still applies:
multiplying and
averaging following Eq.~\eqref{eq:char-filtering-process},
we extract \(p_{\ydiagram{2}}\) and compute \(F(\noise)\).

Therefore,
using a weak coherent state and an intensity measurement,
we obtain a plausible route to estimating the figure of merit in this extended
setting.
Moreover, the form of the data analysis remains unchanged under the loss and
gain model considered here, which simplifies the interpretation of the filtered
signal.

An experimental implementation can be realized by preparing states with
attenuated lasers and measuring them using single-photon avalanche diode click
detection. This method is selected because attenuated lasers represent the
standard, straightforward alternative to ideal single-photon sources in
practical quantum optics~\cite{Scarani2009}. Moreover, single-photon avalanche
diode click detection is considerably simpler and more technologically mature
than photon-number-resolving detectors~\cite{Hadfield2009}.

We now illustrate the preceding discussion with a numerical test of the
fidelity estimate in the presence of loss and gain.
We work with an extended two-mode Hilbert space allowing zero, one, and two photons:
\begin{equation}
  \hilbert_\ext \coloneqq \hilbert^{0}_2 \oplus \hilbert^{1}_2 \oplus \hilbert^{2}_2.
\end{equation}
Here,
\(\hilbert^{n}_2\) denotes the \(n\)-photon subspace for \(m=2\) modes.
For concreteness,
\begin{subequations}
\begin{multline}
\hilbert^{1}_2 \coloneqq 
\ydiagram{1}
=
\operatorname{span}\{
  \GTket{1}{0}{0},
  \GTket{1}{0}{1}
  \}
\\\cong\operatorname{span}
\{\ket{(1,0)},\ket{(0,1)}\}
,
\end{multline}
\begin{equation}
\hilbert^{0}_2 \coloneqq 
\nullirrep
=\operatorname{span}\{\GTket{0}{0}{0}\}
\cong\operatorname{span}\{\ket{(0,0)}\}
,
\end{equation}
and
\begin{multline}
\hilbert^{2}_2 \coloneqq 
\ydiagram{2}
=
\operatorname{span}\{
  \GTket{2}{0}{0},
  \GTket{2}{0}{1},
  \GTket{2}{0}{2}
  \}
\\\cong\operatorname{span}
\{
  \ket{(2,0)},
  \ket{(1,1)},
  \ket{(0,2)}
\}
.
\end{multline}
\end{subequations}
Thus, the decomposition of \(\hilbert_\ext\) into SU(2) irreps is
\begin{multline}\label{eq:decomposition-h-ext-two-photons}
\setchannels(\hilbert_\ext)
\cong
\hilbert_\ext\otimes
\hilbert_\ext^{*}
\\\cong
3\nullirrep +
4 \ydiagram{1}+
4 \ydiagram{2}+
2 \ydiagram{3}+
\ydiagram{4},
\end{multline}
where a number multiplying an irrep means that the irrep appears that number of
times.

The ideal system is a single photon and a two-mode interferometer, so adding
the zero- and two-photon sectors allows us to study loss and gain.
The process of loss and gain is modeled as a \gls{cptp} map~\(\noise\) acting
on \(\hilbert_\ext\).
The goal is to estimate the \gls{agf}
of \(\noise\) restricted to \(\hilbert^{1}_2\).
Moreover, we restrict the measurement to an intensity measurement,
that is,
the data are of the form
\begin{equation}
  d^{(\ngates,s)} =
  \tr\!\left[
    \Pi_{\mathrm{num}} \bigcirc_{U\in \mathbf{U}^{\ngates}_s}\widetilde{\mathcal{U}}(\varrho)
  \right],
\end{equation}
where \(\widetilde{\mathcal{U}}=\noise\circ\mathcal{U}\) (as in Eq.~\eqref{eq:noise-def})
and
\(\Pi_{\mathrm{num}}\)
is the truncated number operator on a single output port:
\begin{equation}
  \Pi_{\mathrm{num}} \coloneqq \dyad{(1,0)} + \dyad{(2,0)}.
\end{equation}
\begin{figure}[t]
  \includegraphics[]{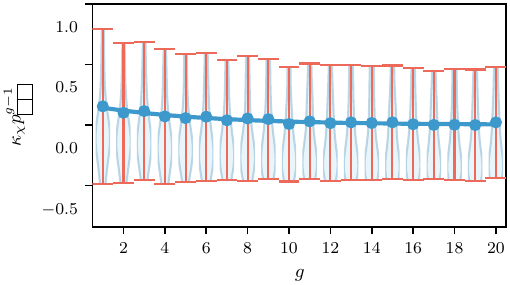}
\caption{\label{fig:violin}Violin plots (with error bars) of the empirical distribution of
\(\koeff_\chi p_{\ydiagram{2}}^{\ngates-1}\) across the sampled circuits for each
depth~\(\ngates\); dots show the mean, and the vertical bars show one standard
deviation, with a smooth fit given by \(\approx 0.17\,(0.83)^{\ngates}\) for
the expression on the vertical axis.}
\end{figure}

The simulation compares the fidelity of a channel on the single-photon
subspace with the fidelity estimated using the single-photon filter.
The channel acts on \(\hilbert_\ext\).
We fix the fidelity over
\(\hilbert_\ext\);
the fixed fidelity channels are randomly sampled according to a known procedure~\cite{Kukulski2021}.
The simulation data are obtained by preparing \(\ket{(1,0)}\), applying a
random unitary, and then performing the intensity measurement defined by
\(\Pi_{\mathrm{num}}\).

To estimate the fidelity on the single-photon Hilbert
space \(\hilbert^{1}_2\), we use the same expression as for qubit systems:
\begin{equation}\label{eq:fidelity-qubit}
F_\text{est}(\noise)
=
6^{-1}(p_{(1,1)} + 3p_{(2,0)})
+3^{-1},
\end{equation}
where the quantities \(p_\mu\) (defined in Eq.~\eqref{eq:parameters})
are used to estimate the fidelity in Eq.~\eqref{eq:representation-gamma}.
The parameters are obtained by using the characters
for the irreps~\(\mu=(1,1)\)
and \(\mu=(2,0)\) of SU(\(2\)):
\(\chi_{(1,1)}(U) = 1\)
and
(see Eq.~\eqref{eq:useful-char})
\begin{equation}
  \chi_{(2,0)}(U) =
  U_{11}^2 + U_{22}^2 + U_{12}U_{21} + U_{11}U_{22}.
\end{equation}
We then compare \(F_\text{est}(\noise)\)
with \(F(\noise\restriction_{\hilbert^{1}_2})\),
where \(\noise\restriction_{\hilbert^{1}_2}\) denotes the restriction of the
channel \(\noise\) to the subspace~\(\hilbert^{1}_2\); that is, the fidelity
restricted to the single-photon subspace.

\begin{figure}
\includegraphics{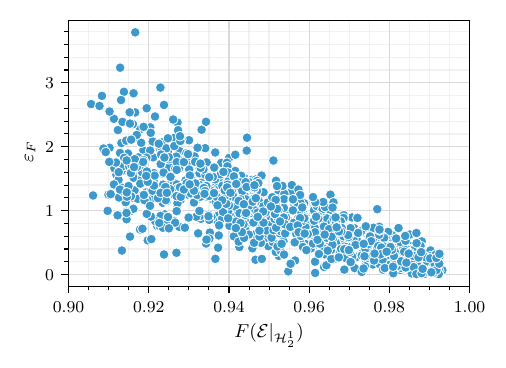}
\caption{\label{fig:fidelity_vs_percentage_error}
  Plot showing the fidelity versus the percentage error of
  Eq.~\eqref{eq:percentage-error} for the fit using our scheme for the gain and loss study. 
  The plot shows \(10^{3}\) points,
  which correspond to~\(10^{3}\) simulations done with different randomly
  sampled channels.
}
\end{figure}

We compare the fidelities (estimated and restricted) in
Fig.~\ref{fig:fidelity_vs_percentage_error}.
The plot contains \(2\times10^{3}\) points;
each point is obtained from \(10^{3}\) so-called shots with
depth up to~20;
that is, for each depth, we sample \(10^{3}\) circuits.
We quantify the discrepancy by the percentage error
\begin{equation}\label{eq:percentage-error}
  \varepsilon_F \coloneqq
  100
  \left(F_\text{est}(\noise)-F(\noise\restriction_{\hilbert^{1}_2})\right)
  F(\noise\restriction_{\hilbert^{1}_2})^{-1}.
\end{equation}
From Fig.~\ref{fig:fidelity_vs_percentage_error}, we see that the method
is able to estimate the fidelity within a \(2\%\)
error
for fidelity values above~\(0.95\),
 and that it overestimates the fidelity;
the estimate improves as the fidelity of
\(\noise\) increases.
The overestimation can be explained by the decomposition in
Eq.~\eqref{eq:decomposition-h-ext-two-photons}.
Whereas~\(\setchannels(\hilbert^{1}_2) \cong \nullirrep + \ydiagram{2}\),
both \(\nullirrep\) and~\(\ydiagram{2}\) appear more than once in
Eq.~\eqref{eq:decomposition-h-ext-two-photons}.
Thus, the filter collects contributions from these additional irreps.

Overall,
these numerical results show that the immanant- and character-based filters
still provide a close approximation to the fidelity even in the presence of gain
and loss errors.
The quantity \(\koeff_\chi\) is, in general, complex.
Thus, fitting an exponential decay requires applying the modulus to the
filtered quantity before performing the fit.
Since the complex prefactor is constant, taking the modulus does not alter the
parameter being estimated.

\subsection{Comparison}

We summarize a comparison between the existing scheme and ours
in Table~\ref{tab:summary-comparison}.
The main difference is that our scheme does not require a projector in the filter definition.
By avoiding the use of a projector,
the data analysis is simplified in two ways:
the number of algebraic terms is reduced, and the computation of
\gls{cg} coefficients is eliminated.
We emphasize that, despite having different filter functions, the same
experimental data can be used in our scheme to obtain the parameters in the
fidelity-like quantity.

Compared to the original filter \(f_{\lambda,\text{orig}}\) of
Eq.~\eqref{eq:eq:original-filter}, our reformulation avoids using the
auxiliary operator \(S\) and its pseudo-inverse \(S^{+}\). This simplifies the computation
of the filter function but does not have a major impact on overall cost. These
changes do not always mean the immanant filter will run faster. Having fewer
immanants does not always lower the total cost, since evaluating immanants of
shape \(\lambda_k\) can be as hard as computing permanents of \(m\times m\) matrices.
The immanant
filter works best when \(n\) is close to \(m\).

We discuss two important cases. For efficient post-processing across many irreps
or large mode numbers, the character filter is preferred because it has
polynomial computational cost and maintains a small, constant variance, making
it scalable as problems grow larger. In contrast, while the immanant filter
simplifies the original method by removing \gls{cg} coefficients, skipping the
auxiliary operator \(S\), and reducing the number of algebraic terms, its overall
computational cost can still be high if the immanants themselves are complex to
evaluate. Our reformulation specifically reduces the overhead of tracking basis
changes and the amount of pre-computation, but does not necessarily reduce the
evaluation cost of the immanants in challenging cases.

\begin{table*}[t]
  \caption{\label{tab:summary-comparison}
    Comparison of the original, immanant, and character filters. The table
    separates algebraic overhead from evaluation cost and indicates the regime
    in which each filter is most useful.
}
  \centering
  \scriptsize
  \begin{tabular}{@{}lllllll@{}}
    \toprule
    Method    & Immanants                & Permanents                             & Filter cost                                                                                                           & \gls{cg} & Filter variance & Preferred regime\\ \midrule
    Original  & 1                        & \(\sharp_{\lambda} - 1 + d_{\lambda}\) & \(\geq O(2^{n} n^2)\)\footnote{This does not consider the cost of looking up or computing the \gls{cg} coefficients.} & Yes      & \shortstack[l]{Estimated\\ \(\log\)-like~\cite{mirko2025}} & \shortstack[l]{Useful baseline when \(n\ll m\),\\ since its dominant permanents are\\ \(n\times n\), although it retains\\ projector and \gls{cg} overhead.}\\
    Immanant  & \(\sharp_{\lambda} - 1\) & 1
              & \(O(2^{m} m^2)\)\footnote{Worst case.}                                                                                & No       & \shortstack[l]{Upper-bounded by\\ \(1/\binom{m}{n}\)~\cite{Daigle2025}} & \shortstack[l]{Most useful when \(n\approx m\)\\ and reducing projector/\gls{cg}\\ overhead matters more than\\ worst-case asymptotic.}\\
    Character & 0                        & 0                                      & Polynomial~\cite{Buergisser2000a}                                                                                     & No       & 1                & \shortstack[l]{Preferred for large-scale\\ post-processing, many irreps,\\ or large mode number, since it\\ avoids immanants altogether.}\\\bottomrule
  \end{tabular}
\end{table*}

Two limiting regimes appear in Table~\ref{tab:summary-comparison}.
If \(n\ll m\), the original scheme may still have a lower computation cost
because it is dominated by \(n\times n\) permanents, whereas the immanant
filter involves immanants of size \(m\). If \(n\approx m\), the immanant
filter becomes more attractive because it removes the \gls{cg} coefficients
and reduces the number of algebraic terms that must be assembled. In either
regime, the character filter remains the efficient option, as
it avoids both permanents and immanants and has polynomial evaluation cost.

We first comment on the number of algebraic objects required in the immanant
filter and compare it with the original scheme.
Let \( \sharp_{\lambda} \) denote the number of irreps
in the decomposition of \(\Gamma\). 
Let \( d_{\lambda} \) denote the dimension of~\(\Gamma_\lambda\).  
Then, the lower bound on the number of permanents needed is  
\begin{equation}
\sharp_{\lambda} - 1 + d_{\lambda}.
\end{equation}  
On the other hand,
the number of immanants is
\begin{equation}
\sharp_{\lambda} - 1.
\end{equation}
Therefore, the number of permanents using the original scheme in
Ref.~\cite{mirko2025}
is strictly larger than the number of immanants required by our immanant filter.
Note that the bound is not tight,
but it suffices to show that our reformulation reduces the number of algebraic
terms.
This comparison concerns the number of terms only; it does not by itself
resolve the cost of evaluating each immanant.
The most significant simplification common to both of our filters is the
elimination of the \gls{cg} coefficients.
In addition, the character filter requires neither permanents nor immanants
and, by Remarks~\ref{rem:char-variance} and~\ref{rem:efficient}, has
polynomial evaluation cost and constant variance.

The main trade-off lies with the immanant filter. In the original approach, the
filter uses permanents, as seen in Eq. (16) of Ref.~\cite{mirko2025}. 
By contrast, our immanant filter
replaces projector-dependent permanents with fewer immanants and eliminates
\gls{cg}
coefficients; this reduces algebraic overhead, but immanants can still be
expensive to compute. By contrast, the character filter in Sec.~\ref{sec:approach} avoids
immanant evaluation and allows efficient computation for any irrep. Immanants
have more complicated variance~\cite{Daigle2025,DaigleGuiseWelshInPrep},
while the character filter’s variance is~1.
For these reasons, the character filter emerges as the most efficient of the three in both runtime and sampling cost.

\section{Discussion and conclusion}
\label{sec:discussion}

In this section,
we summarize the main limitations of the state-of-the-art and how our scheme
addresses them.
We emphasize the advantages in both data analysis and experimental
implementation, and we outline directions for future work.

On the data-analysis side,
the original approach requires many matrix permanents,
each determined via \gls{cg} coefficient expansions.
This renders the procedure computationally demanding and analytically opaque.
Our reformulation removes the \gls{cg}-coefficient overhead and reduces the
number of algebraic terms entering the filter.
The immanant filter replaces most of the original permanent expressions by a
smaller set of immanants~\cite{Buergisser2000b}, while the character filter
replaces them altogether by SU(\(m\)) characters.
This makes the filtering formulas more transparent and easier to implement in
standard computer algebra systems.

These improvements should be interpreted with care. The immanant filter uses
fewer algebraic objects, but evaluating the relevant immanants in the
decomposition of \(\Gamma\) in
Eq.~\eqref{eq:irrep-decomposition-gamma} remains computationally expensive and is closer in
cost to permanents than to determinants~\cite{Buergisser2000a}. The immanant filter is preferable
when \(n\) is comparable to \(m\), as the original scheme is dominated by permanents. By
contrast, the character filter avoids immanant evaluation, operates in
polynomial time, and has known constant variance, making it the genuinely
efficient option among the filters considered.

On the experimental side,
the standard protocol assumes photon-number-resolving detectors and multiple
Fock-state preparations, which are challenging for many platforms.
Our numerical and theoretical analysis suggests that these assumptions can be relaxed while
still obtaining a reasonable estimate of the target figure of merit.
This points to a feasible route based on weak coherent states and
coarse-grained (intensity) detection.

Despite using different filters,
our scheme retains the core feature of the original method:
it produces a single-exponential decay in the benchmarking signal from which
the figure of merit can be computed.
Our immanant and character schemes offer some advantages: both fewer and less
complex algebraic objects, and these can be evaluated using tools such as
\texttt{GAP} or \texttt{Wolfram}, without specialized routines for SU(\(m\))
\gls{cg} coefficients.
Our main contribution is to show that D-functions in irreps already contain a
projector. Using them as filters (immanants or characters) removes the need for
explicit projectors and basis changes. However, the filters differ in practice.
Immanant filters reduce algebraic effort but may still be costly due to the
matrices required. Character filters work the same way but avoid these costs.
Thus, with Remarks~\ref{rem:char-variance} and~\ref{rem:efficient}, character filtering is the most practical option.

Overall,
by reducing both the computational and experimental demands of the original benchmarking method,
our scheme greatly improves the practical application of benchmarking passive bosonic channels.
Looking into future work,
an intriguing avenue for future research is expanding this framework to include active bosonic transformations.
However,
such generalizations are challenging,
primarily due to the non-compactness of the relevant transformation group
and difficulties with the generalization of the concept of fidelity.
Nevertheless,
our scheme’s simple handling of loss and gain errors provides a strong foundation for practical benchmarking in continuous-variable quantum technologies.

\section*{Data and code availability}
The data and source code used in this work are available at Zenodo:
\href{https://doi.org/10.5281/zenodo.19493886}{10.5281/zenodo.19493886}.

\section*{Acknowledgments}
The author is grateful to Dr.\ Hubert de~Guise
and Prof.\ Dr.\ Stefan Scheel for helpful discussions.
The author also acknowledges
help from
Dr.\ M\'ario Ziman,
Dr.\ Seyed Arash Ghoreishi,
and
Dr.\ Konrad Szyma\'nski
in the preparation of the
manuscript.
The author also acknowledges support from
projects
DeQHOST APVV-22-0570,
and
QUAS VEGA 2/0164/25.

\bibliography{references}

@PREAMBLE{
 "\providecommand{\noopsort}[1]{}" 
 # "\providecommand{\singleletter}[1]{#1}%" 
}

@article{proctor2017,
  title        = {{What randomized benchmarking actually measures}},
  author       = {Proctor, Timothy and Rudinger, Kenneth and Young, Kevin and others},
  year         = 2017,
  journal      = {Phys. Rev. Lett.},
  publisher    = {American Physical Society},
  volume       = 119,
  number       = 13,
  pages        = 130502,
  doi          = {10.1103/PhysRevLett.119.130502}
}

@article{Alex2011,
  title = {A numerical algorithm for the explicit calculation of {SU(N)} and {$\mbox{SL}(N, \mathbb {C})$} {Clebsch–Gordan} coefficients},
  volume = {52},
  ISSN = {1089-7658},
  url = {http://dx.doi.org/10.1063/1.3521562},
  DOI = {10.1063/1.3521562},
  number = {2},
  journal = {J. Phys. A: Math. Theor.},
  publisher = {AIP Publishing},
  author = {Alex,  Arne and Kalus,  Matthias and Huckleberry,  Alan and von Delft,  Jan},
  year = {2011},
  month = feb 
}

@article{mirko2025,
  title = {Bosonic Randomized Benchmarking with Passive Transformations},
  author = {Arienzo, Mirko and Grinko, Dmitry and Kliesch, Martin and Heinrich, Markus},
  journal = {PRX Quantum},
  volume = {6},
  issue = {2},
  pages = {020305},
  numpages = {41},
  year = {2025},
  month = {Apr},
  publisher = {American Physical Society},
  doi = {10.1103/PRXQuantum.6.020305},
  url = {https://link.aps.org/doi/10.1103/PRXQuantum.6.020305}
}

@article{Kostan1995,
  title = {Immanant Inequalities and 0-Weight Spaces},
  volume = {8},
  ISSN = {0894-0347},
  url = {http://dx.doi.org/10.2307/2152885},
  DOI = {10.2307/2152885},
  number = {1},
  journal = {J. Am. Math. Soc.},
  publisher = {JSTOR},
  author = {Kostant,  Bertram},
  year = {1995},
  month = jan,
  pages = {181}
}

@article{deguise2018,
  title = {Simple factorization of unitary transformations},
  author = {de Guise, Hubert and Di Matteo, Olivia and S\'anchez-Soto, Luis L.},
  journal = {Phys. Rev. A},
  volume = {97},
  issue = {2},
  pages = {022328},
  numpages = {7},
  year = {2018},
  month = {Feb},
  publisher = {American Physical Society},
  doi = {10.1103/PhysRevA.97.022328},
  url = {https://link.aps.org/doi/10.1103/PhysRevA.97.022328}
}

@BOOK{littlewood1977,
  title     = {The Theory of Group Characters and Matrix Representations of
               Groups},
  author    = "Littlewood, Dudley Ernest",
  publisher = "American Mathematical",
  year      =  1977,
}

@article{Guise2016,
  title = {D-functions and immanants of unitary matrices and submatrices},
  volume = {49},
  ISSN = {1751-8121},
  url = {http://dx.doi.org/10.1088/1751-8113/49/9/09LT01},
  DOI = {10.1088/1751-8113/49/9/09lt01},
  number = {9},
  journal = {J. Phys. A: Math. Theor.},
  publisher = {IOP Publishing},
  author = {de Guise,  Hubert  and Spivak,  Dylan and Kulp,  Justin and Dhand,  Ish},
  year = {2016},
  month = jan,
  pages = {09LT01}
}

@article{Masada2015,
  title = {Continuous-variable entanglement on a chip},
  volume = {9},
  ISSN = {1749-4893},
  url = {http://dx.doi.org/10.1038/nphoton.2015.42},
  DOI = {10.1038/nphoton.2015.42},
  number = {5},
  journal = {Nat. Photonics},
  publisher = {Springer Science and Business Media LLC},
  author = {Masada,  Genta and Miyata,  Kazunori and Politi,  Alberto and Hashimoto,  Toshikazu and O’Brien,  Jeremy L. and Furusawa,  Akira},
  year = {2015},
  month = mar,
  pages = {316–319}
}

@article{Takeda2017,
  title = {Universal Quantum Computing with Measurement-Induced Continuous-Variable Gate Sequence in a Loop-Based Architecture},
  author = {Takeda, Shuntaro and Furusawa, Akira},
  journal = {Phys. Rev. Lett.},
  volume = {119},
  issue = {12},
  pages = {120504},
  numpages = {5},
  year = {2017},
  month = {Sep},
  publisher = {American Physical Society},
  doi = {10.1103/PhysRevLett.119.120504},
  url = {https://link.aps.org/doi/10.1103/PhysRevLett.119.120504}
}

@article{Yonezu2023,
  title = {Time-Domain Universal Linear-Optical Operations for Universal Quantum Information Processing},
  author = {Yonezu, Kazuma and Enomoto, Yutaro and Yoshida, Takato and Takeda, Shuntaro},
  journal = {Phys. Rev. Lett.},
  volume = {131},
  issue = {4},
  pages = {040601},
  numpages = {6},
  year = {2023},
  month = {Jul},
  publisher = {American Physical Society},
  doi = {10.1103/PhysRevLett.131.040601},
  url = {https://link.aps.org/doi/10.1103/PhysRevLett.131.040601}
}

@article{Fukui2022,
  title = {Building a large-scale quantum computer with continuous-variable optical technologies},
  volume = {55},
  ISSN = {1361-6455},
  url = {http://dx.doi.org/10.1088/1361-6455/ac489c},
  DOI = {10.1088/1361-6455/ac489c},
  number = {1},
  journal = {J. Phys. B: At. Mol. Opt. Phys.},
  publisher = {IOP Publishing},
  author = {Fukui,  Kosuke and Takeda,  Shuntaro},
  year = {2022},
  month = jan,
  pages = {012001}
}

@article{Buergisser2000b,
  title = {The Computational Complexity of Immanants},
  volume = {30},
  ISSN = {1095-7111},
  url = {http://dx.doi.org/10.1137/S0097539798367880},
  DOI = {10.1137/s0097539798367880},
  number = {3},
  journal = {SIAM J. Comput.},
  publisher = {Society for Industrial \& Applied Mathematics (SIAM)},
  author = {B\"{u}rgisser,  Peter},
  year = {2000},
  month = jan,
  pages = {1023–1040}
}

@article{Buergisser2000a,
  author  = {B{\"u}rgisser, Peter},
  title   = {The Computational Complexity to Evaluate Representations of General Linear Groups},
  journal = {SIAM Journal on Computing},
  volume  = {30},
  number  = {3},
  pages   = {1010--1022},
  year    = {2000},
  doi     = {10.1137/S0097539798367892}
}

@article{Valiant1979,
  title = {The complexity of computing the permanent},
  volume = {8},
  ISSN = {0304-3975},
  url = {http://dx.doi.org/10.1016/0304-3975(79)90044-6},
  DOI = {10.1016/0304-3975(79)90044-6},
  number = {2},
  journal = {Theor. Comput. Sci.},
  publisher = {Elsevier BV},
  author = {Valiant,  L.G.},
  year = {1979},
  pages = {189–201}
}

@misc{wilkens2024,
      title={Benchmarking bosonic and fermionic dynamics}, 
      author={Jadwiga Wilkens and Marios Ioannou and Ellen Derbyshire and Jens Eisert and Dominik Hangleiter and Ingo Roth and Jonas Haferkamp},
      year={2024},
      eprint={2408.11105},
      archivePrefix={arXiv},
      primaryClass={quant-ph},
      url={https://arxiv.org/abs/2408.11105}, 
}

@article{Emerson2005,
  title = {Scalable noise estimation with random unitary operators},
  volume = {7},
  ISSN = {1741-3575},
  url = {http://dx.doi.org/10.1088/1464-4266/7/10/021},
  DOI = {10.1088/1464-4266/7/10/021},
  number = {10},
  journal = {J. Opt. B: Quantum Semiclassical Opt.},
  publisher = {IOP Publishing},
  author = {Emerson,  Joseph and Alicki,  Robert and Życzkowski,  Karol},
  year = {2005},
  month = sep,
  pages = {S347–S352}
}

@article{Magesan2011,
  title = {Scalable and Robust Randomized Benchmarking of Quantum Processes},
  volume = {106},
  ISSN = {1079-7114},
  url = {http://dx.doi.org/10.1103/PhysRevLett.106.180504},
  DOI = {10.1103/physrevlett.106.180504},
  number = {18},
  journal = {Phys. Rev. Lett.},
  publisher = {American Physical Society (APS)},
  author = {Magesan,  Easwar and Gambetta,  J. M. and Emerson,  Joseph},
  year = {2011},
  month = may 
}

@article{Knill2008,
  title = {Randomized benchmarking of quantum gates},
  volume = {77},
  ISSN = {1094-1622},
  url = {http://dx.doi.org/10.1103/PhysRevA.77.012307},
  DOI = {10.1103/physreva.77.012307},
  number = {1},
  journal = {Phys. Rev. A},
  publisher = {American Physical Society (APS)},
  author = {Knill,  E. and Leibfried,  D. and Reichle,  R. and Britton,  J. and Blakestad,  R. B. and Jost,  J. D. and Langer,  C. and Ozeri,  R. and Seidelin,  S. and Wineland,  D. J.},
  year = {2008},
  month = jan 
}

@article{AmaroAlcal2024,
  title = {Randomised benchmarking for universal qudit gates},
  volume = {26},
  ISSN = {1367-2630},
  url = {http://dx.doi.org/10.1088/1367-2630/ad6635},
  DOI = {10.1088/1367-2630/ad6635},
  number = {7},
  journal = {New J. Phys.},
  publisher = {IOP Publishing},
  author = {Amaro-Alcalá,  David and Sanders,  Barry C and de Guise,  Hubert},
  year = {2024},
  month = jul,
  pages = {073052}
}

@article{Jafarzadeh2020,
  title = {{Randomized benchmarking for qudit Clifford gates}},
  volume = {22},
  ISSN = {1367-2630},
  url = {http://dx.doi.org/10.1088/1367-2630/ab8ab1},
  DOI = {10.1088/1367-2630/ab8ab1},
  number = {6},
  journal = {New J. Phys.},
  publisher = {IOP Publishing},
  author = {Jafarzadeh,  Mahnaz and Wu,  Ya-Dong and Sanders,  Yuval R and Sanders,  Barry C},
  year = {2020},
  month = jun,
  pages = {063014}
}

@article{helsen2022,
  title = {General Framework for Randomized Benchmarking},
  author = {Helsen, J. and Roth, I. and Onorati, E. and Werner, A.H. and Eisert, J.},
  journal = {PRX Quantum},
  volume = {3},
  issue = {2},
  pages = {020357},
  numpages = {54},
  year = {2022},
  month = {Jun},
  publisher = {American Physical Society},
  doi = {10.1103/PRXQuantum.3.020357},
  url = {https://link.aps.org/doi/10.1103/PRXQuantum.3.020357}
}

@article{want2019,
  title = {{Boson Sampling with 20 Input Photons and a 60-Mode Interferometer in a $1{0}^{14}$-Dimensional Hilbert Space}},
  author = {Wang, Hui and Qin, Jian and Ding, Xing and Chen, Ming-Cheng and Chen, Si and You, Xiang and He, Yu-Ming and Jiang, Xiao and You, L. and Wang, Z. and Schneider, C. and Renema, Jelmer J. and H\"ofling, Sven and Lu, Chao-Yang and Pan, Jian-Wei},
  journal = {Phys. Rev. Lett.},
  volume = {123},
  issue = {25},
  pages = {250503},
  numpages = {7},
  year = {2019},
  month = {Dec},
  publisher = {American Physical Society},
  doi = {10.1103/PhysRevLett.123.250503},
  url = {https://link.aps.org/doi/10.1103/PhysRevLett.123.250503}
}

@book{raczka1986,
  title={{Theory of Group Representations and Applications}},
  author={Raczka, Ryszard and Barut, Asim Orhan},
  year={1986},
  publisher={World Scientific Publishing Company}
}

@book{Fulton1996,
  title = {Young Tableaux: With Applications to Representation Theory and Geometry},
  ISBN = {9780511626241},
  url = {http://dx.doi.org/10.1017/CBO9780511626241},
  DOI = {10.1017/cbo9780511626241},
  publisher = {Cambridge University Press},
  author = {Fulton,  William},
  year = {1996},
  month = dec 
}

@misc{WolframImmanant,
  author       = {Wolfram},
  title        = {Immanant},
  howpublished = {\url{https://resources.wolframcloud.com/FunctionRepository/resources/Immanant/}},
  note         = {Accessed: 2025-10-07}
}

@article{wallman2016,
  title = {Noise tailoring for scalable quantum computation via randomized compiling},
  author = {Wallman, Joel J. and Emerson, Joseph},
  journal = {Phys. Rev. A},
  volume = {94},
  issue = {5},
  pages = {052325},
  numpages = {9},
  year = {2016},
  month = {Nov},
  publisher = {American Physical Society},
  doi = {10.1103/PhysRevA.94.052325},
  url = {https://link.aps.org/doi/10.1103/PhysRevA.94.052325}
}

@article{Hashim2021,
  title = {Randomized Compiling for Scalable Quantum Computing on a Noisy Superconducting Quantum Processor},
  author = {Hashim, Akel and Naik, Ravi K. and Morvan, Alexis and Ville, Jean-Loup and Mitchell, Bradley and Kreikebaum, John Mark and Davis, Marc and Smith, Ethan and Iancu, Costin and O'Brien, Kevin P. and Hincks, Ian and Wallman, Joel J. and Emerson, Joseph and Siddiqi, Irfan},
  journal = {Phys. Rev. X},
  volume = {11},
  issue = {4},
  pages = {041039},
  numpages = {12},
  year = {2021},
  month = {Nov},
  publisher = {American Physical Society},
  doi = {10.1103/PhysRevX.11.041039},
  url = {https://link.aps.org/doi/10.1103/PhysRevX.11.041039}
}

@article{AmaroAlcal2020,
  title = {Sum rules in multiphoton coincidence rates},
  volume = {384},
  ISSN = {0375-9601},
  url = {http://dx.doi.org/10.1016/j.physleta.2020.126459},
  DOI = {10.1016/j.physleta.2020.126459},
  number = {20},
  journal = {Phys. Lett. A},
  publisher = {Elsevier BV},
  author = {Amaro-Alcalá,  David and Spivak,  Dylan and de Guise,  Hubert},
  year = {2020},
  month = jul,
  pages = {126459}
}

@article{Lin2019,
  title = {On the freedom in representing quantum operations},
  volume = {21},
  ISSN = {1367-2630},
  url = {http://dx.doi.org/10.1088/1367-2630/ab075a},
  DOI = {10.1088/1367-2630/ab075a},
  number = {2},
  journal = {New J. Phys.},
  publisher = {IOP Publishing},
  author = {Lin,  Junan and Buonacorsi,  Brandon and Laflamme,  Raymond and Wallman,  Joel J},
  year = {2019},
  month = feb,
  pages = {023006}
}

@article{Daigle2025,
  title = {Mixed symmetries of $\mathfrak{S}_n$: immanants in the sampling of ${U}(d)$ submatrices},
  volume = {3152},
  ISSN = {1742-6596},
  url = {http://dx.doi.org/10.1088/1742-6596/3152/1/012020},
  DOI = {10.1088/1742-6596/3152/1/012020},
  number = {1},
  journal = {J. Phys. Conf. Ser.},
  publisher = {IOP Publishing},
  author = {Daigle, Jacob and {de Guise}, Hubert and Welsh, Trevor},
  year = {2025},
  month = dec,
  pages = {012020}
}

@unpublished{DaigleGuiseWelshInPrep,
  title  = {Mixed symmetries and immanants of $\mathfrak{S}_n$: the next generation},
  author = {Daigle, Jacob and {de Guise}, Hubert and Welsh, Trevor},
  note   = {Manuscript in preparation},
  year   = {2025}
}

@misc{GroupFunctionsjl,
  title        = {{GroupFunctions.jl}},
  author       = {Amaro-Alcalá, David},
  howpublished = {\url{https://github.com/davidamaro/GroupFunctions.jl}},
  year         = {2025}
}

@article{Kukulski2021,
  title = {Generating random quantum channels},
  volume = {62},
  ISSN = {1089-7658},
  url = {http://dx.doi.org/10.1063/5.0038838},
  DOI = {10.1063/5.0038838},
  number = {6},
  journal = {J. Math. Phys.},
  publisher = {AIP Publishing},
  author = {Kukulski, Ryszard and Nechita, Ion and Pawela, {\L}ukasz and Pucha{\l}a, Zbigniew and {\.Z}yczkowski, Karol},
  year = {2021},
  month = jun 
}

@article{Scarani2009,
  title = {The security of practical quantum key distribution},
  author = {Scarani, Valerio and Bechmann-Pasquinucci, Helle and Cerf, Nicolas J. and Du\ifmmode \check{s}\else \v{s}\fi{}ek, Miloslav and L\"utkenhaus, Norbert and Peev, Momtchil},
  journal = {Rev. Mod. Phys.},
  volume = {81},
  issue = {3},
  pages = {1301--1350},
  numpages = {0},
  year = {2009},
  month = {Sep},
  publisher = {American Physical Society},
  doi = {10.1103/RevModPhys.81.1301},
  url = {https://link.aps.org/doi/10.1103/RevModPhys.81.1301}
}

@article{Hadfield2009,
  title = {Single-photon detectors for optical quantum information applications},
  volume = {3},
  ISSN = {1749-4893},
  url = {http://dx.doi.org/10.1038/nphoton.2009.230},
  DOI = {10.1038/nphoton.2009.230},
  number = {12},
  journal = {Nat. Photonics},
  publisher = {Springer Science and Business Media LLC},
  author = {Hadfield,  Robert H.},
  year = {2009},
  month = dec,
  pages = {696–705}
}

\appendix
\section{Example Kostant's relation for SU(3)}
\label{app:kostant-relation-example}
In this appendix,
we illustrate several cases of the general result,
labeled Kostant’s relation,
stated in Theorem~\ref{thm:kostant}.
To achieve this comparison,
we need to compute D-functions for SU(3),
and then compute the immanants using the relations in Eq.~\eqref{eq:imm-novel}.
We then verify that both yield the same result.

The second out of three ingredients is the zero-weight states.
Applying the formula in Eq.~\eqref{eq:weight},
we obtain that the zero-weight states for \(\mu = (2,1,0)\) are 
\begin{equation}\label{eq:app-zw-df}
  \ket{\zeta_{(2,1,0)}^{(0)}}=
\Ket{\footnotesize
  \setlength{\arraycolsep}{1pt}
  \begin{array}{ccccccc}
2& &1& &0 \\
 &1& &1&  \\
 & &1& &  
\end{array}}
\text{and}
  \ket{\zeta_{(2,1,0)}^{(1)}} =
  \Ket{\footnotesize
  \setlength{\arraycolsep}{1pt}
\begin{array}{ccccccc}
2& &1& &0\\
 &2& &0& \\
 & &1& & \\
\end{array}}.
\end{equation}
We then use the patterns~\eqref{eq:app-zw-df} to compute immanants.
To compute D-functions for SU(2) and SU(3),
we utilize the results listed in Ref.~\cite{Alex2011}.
The formulas for the generalized rising and lowering operators 
are presented in Eqs.~(28) and~(29) therein.
Then,
using the simple factorization of unitary operations (Ref.~\cite{deguise2018}),
we can compute the representations of SU(2) and SU(3) matrices for any partition \(\mu\).
For conciseness, we pick \(\mu = (2,1,0)\) for SU(3).
The diagonal entries for the irrep \(\mu\) of SU(3) corresponding to the
zero-weight states, the set \(\mathcal{Z}_\mu\), in~Eq.~\eqref{eq:app-zw-df}
are
\begin{subequations}\label{eq:subeq-zero-weight-states-example}
\begin{align}
&\Bra{
  {
  \setlength{\arraycolsep}{1pt}
  \footnotesize
\begin{array}{ccccccc}
2& &1& &0 \\
 &1& &1&  \\
 & &1& &  
\end{array}
}
}\Gamma_{\mu = (2,1,0)}(U)\Ket{
  \setlength{\arraycolsep}{1pt}
\footnotesize\begin{array}{ccccccc}
2& &1& &0 \\
 &1& &1&  \\
 & &1& &  
\end{array}
}
\label{eq:app-entry-a}\\&=
\nicefrac{1}{16}
\begin{aligned}[t]
\Big(
  &1 - 3\cos\beta_2(\cos\beta_3-1)\\
  & + 3\cos\beta_3 + 3 \cos\beta_1\\
  &\quad(1 + \cos\beta_2(\cos\beta_3-1) + 3\cos\beta_3)\\
  &-12\cos\nicefrac{\beta_2}{2}\cos(\alpha_2-\alpha_3-\gamma_1)\sin\beta_1\sin\beta_3\Big)
\end{aligned}
\nonumber
\end{align}
\end{subequations}
and
\begin{align}
&\Bra{
  {
  \setlength{\arraycolsep}{1pt}
\footnotesize\begin{array}{ccccccc}
2& &1& &0 \\
 &2& &0&  \\
 & &1& &  
\end{array}
}
}\Gamma_{\mu = (2,1,0)}(U)\Ket{
  \setlength{\arraycolsep}{1pt}
\footnotesize\begin{array}{ccccccc}
2& &1& &0 \\
 &2& &0&  \\
 & &1& &  
\end{array}
}
\label{eq:app-entry-b}\\&=
\nicefrac{1}{16}\Big(
\begin{aligned}[t]
&-4\sin\beta_1\,\sin\beta_3\,\cos\!\nicefrac{\beta_2}{2}\,
  \cos(\alpha_2-\alpha_3-\gamma_1)\\
&\quad-3\cos\beta_3
+3\cos\beta_2(\cos\beta_3+3)\\
&\quad+\cos\beta_1\big(3(\cos\beta_3-1)
\\&+\cos\beta_2(\cos\beta_3+3)\big)
+3\Big),
\end{aligned}
\nonumber
\end{align}
where the angles \(\alpha\), \(\beta\), and \(\gamma\) represent the parameters
of a SU(3) transformation~\cite{deguise2018}.

We now carry out the comparison.
We compute the trace over the zero-weight states \(\mu = (2,0)\).
Next,
computing the immanant for the fundamental irreps (\(\mu = (1,0,0)\) for SU(3)),
we get:
\begin{equation}\label{eq:app-imm}
\begin{aligned}[t]
  \imm_{(2,1,0)}&(U) = 
\\&\nicefrac{1}{4} \big( \sin \left(\beta _3\right) \cos
   \left(\frac{\beta _2}{2}\right) \cos \left(\alpha _2-\alpha _3-\gamma
   _1\right)\\
   &+3 \cos \left(\beta _1\right) \cos \left(\beta _3\right)\\
   &+\cos \left(\beta _2\right) \left(\cos \left(\beta _1\right) \cos \left(\beta _3\right)+3\right)
 +1\big).
 \end{aligned}
\end{equation}
Adding Eq.~\eqref{eq:app-entry-a} and
\eqref{eq:app-entry-b} 
we get Eq.~\eqref{eq:app-imm},
thus corroborating the Kostant relation in
Theorem~\ref{thm:kostant}.

\section{Decomposition of the tensor product symmetric irrep and its
dual}
\label{app:decomposition}
This appendix is divided into three parts.
First,
we recall the notation for the dual irrep.
Next,
we discuss the decomposition of the tensor product of a symmetric irrep and its dual;
we use a different result than in the original work.
We then conclude with the calculation of parameters for the figure of merit
of a noisy gate using the original filter.

We outline the diagrammatic method for identifying the dual irrep of a given irrep \(\lambda\).
This is a specific case within the general algorithm.
To find the dual irreducible representation from the tableau,
first embed it into an \(m\times m\) grid of unlabeled boxes.
The boxes representing the original partition are labeled lambda.
Below these,
label the boxes by \(\lambda^{*}\).
The final shape corresponds to the label.
Below, we present the case for the SU(3) irrep labeled \((2,0,0)\).
\begin{align}
  \ydiagram{3,3,3}
\to
\begin{ytableau}
  \lambda & \lambda & . \\
  . & . & . \\
  . & . & . 
\end{ytableau}\to
\begin{ytableau}
  \lambda & \lambda & . \\
  \lambda^* & \lambda^* & . \\
  \lambda^* & \lambda^* & . 
\end{ytableau}
\to
\lambda^{*} = \ydiagram{2,2}.
\end{align}

The decomposition
of the representation \(\lambda\otimes\lambda^{*}\)
can be carried out in many ways.
To add a novel approach,
we delineate a solution that is less elaborate than the one offered in the original scheme~\cite{mirko2025}.
We note that  Pieri's formula is a less general result compared to the Richardson-Littlewood
formula;
however, it suffices to describe the reduction.
This formula is employed due to the isomorphism that exists between irreps
of the unitary group and symmetric polynomials~\cite{Fulton1996}.

Within the context of irreps of SU(\(m\)), 
each irrep \(\mu\) is isomorphic to a symmetric polynomial 
\(s_\mu\),
which are not to be confused with the coefficients in
Eq.~\eqref{eq:coefficients-s}.
Pieri's formula states that
 \begin{equation}\label{eq:pieri}
s_\lambda s_{\lambda^{*}}
=
\sum_\mu s_\mu,
\end{equation}
where the sum over \(\mu\) corresponds to the partitions obtained from
\(\lambda\) by adding \(n\) elements in different columns and keeping a valid
Young tableau, or equivalently, a valid partition with non-increasing numbers.

We offer two examples to illustrate Pieri's formula in Eq.~\eqref{eq:pieri}:
one for the case SU(2) and the other for SU(3).
The case for SU(2) shows a result that
angular momentum rules
can also obtain.
The first case is that of the partition \(\lambda = (1,0)\).
In that case,
the dual is \(\lambda^{*} = (1,0)\).
There are two ways to add a box to the diagram \(\ydiagram{1}\):
\begin{equation}
\ydiagram{1,1}, \ydiagram{2,0}.
\end{equation}
since a single box for SU(2) denotes a spin 1/2 particle,
the Hilbert space of two spin 1/2 particles decomposes into a spin-zero and a
spin-one subsystems.

The less familiar case arises from considering the irrep \(\lambda = (2,0,0)\)
of SU(3).
The dual, obtained by the process explained at the beginning of this appendix,
is~\(\lambda^{*} = (2,2,0)\).
Thus, to compute the decomposition from Pieri's formula, 
we obtain the different ways we can add two boxes to \(\lambda^{*}\).
These are:
\begin{equation}
\ydiagram{2,2,2},
\ydiagram{3,2,1},
\ydiagram{4,2}.
\end{equation}
Thus,
by simply adding boxes,
the elements in the decomposition in Eq.~\eqref{eq:irrep-decomposition-gamma} can be computed.

We conclude this appendix,
showing how the original filter can be used to obtain the parameters of the
noise~\(\noise\).
We begin by computing a constant for the case where a single gate is present.
Then we describe the procedure for  \(g+1\) gates.

The filter for bosonic \gls{rb} is
\begin{equation}
  f_{\mu,\text{orig}}
  \coloneqq 
  \bra{\varrho} P_\mu S^{+} \Gamma(U)^{\dagger} \ket{E}.
\end{equation}
From \(f_{\mu,\text{orig}}\),
we obtain
\begin{subequations}
  \begin{align}
&\average_U
f_{\mu,\text{orig}}
\bra{\widetilde{E}} \Gamma(U) \ket{\widetilde{\varrho}}
\\&=
\average_U
\bra{\varrho} P_\mu S^{+} \Gamma(U)^{\dagger} \ket{E}
\bra{\widetilde{E}} \Gamma(U) \ket{\widetilde{\varrho}}
\\&=
    \bra{\varrho} P_\mu S^{+} S \ket{\widetilde{\varrho}} \\&\approx
    c_\mu^{\varrho,\widetilde{\varrho}} \coloneqq  \bra{\varrho} P_\mu
    \ket{\widetilde{\varrho}} .
  \end{align}
\end{subequations}

For two gates we have
\begin{subequations}
  \begin{align}
&\average_{U_1,U_0}
\bra{\varrho} P_\mu S^{+}
\Gamma(U_1)^{\dagger}
\Gamma(U_0)^{\dagger}
\ket{E}
\\& \qquad\bra{\widetilde{E}}
\Gamma(U_0)\Gamma(\noise)
\Gamma(U_1)\Gamma(\noise)
\ket{\widetilde{\varrho}}\\
  &=
  \average_{U_1}
  \bra{\varrho} P_\mu S^{+}
  \Gamma(U_1)^{\dagger}
  S
  \Gamma(\noise)
  \Gamma(U_1)\Gamma(\noise)
  \ket{\widetilde{\varrho}}\\
  &=
  \average_{U_1}
  \bra{\varrho} P_\mu S^{+}
  S
  \Gamma(U_1)^{\dagger}
  \Gamma(\noise)
  \Gamma(U_1)\Gamma(\noise)
  \ket{\widetilde{\varrho}}\\
  &\approx
  \average_{U_1}
  \bra{\varrho} P_\mu
  \Gamma(U_1)^{\dagger}
  \Gamma(\noise)
  \Gamma(U_1)\Gamma(\noise)
  \ket{\widetilde{\varrho}}\\
  &=
  \bra{\varrho}
  P_\mu T[\noise]
  \Gamma(\noise)
  \ket{\widetilde{\varrho}}
  =
  \bra{\varrho}
  T_\mu[\noise]
  \Gamma(\noise)
  \ket{\widetilde{\varrho}}
  ,
  \end{align}
\end{subequations}
with
\(T[\noise]\)  
in Eq.~\eqref{eq:twirl-definition}
and
\begin{equation}
T_\mu[\noise]
\coloneqq 
P_\mu T[\noise].
\end{equation}

By Schur's lemma,
\begin{equation}\label{eq:schur-lemma}
T_\mu[\noise] = \param_{\mu, \noise} \mathbb{I}_\mu.
\end{equation}
Thus,
\begin{align}
p_{\mu, \noise} c_\mu^{\varrho, \widetilde{\varrho}}
&=
\average_{U_1,U_0}
\bra{\varrho} P_\mu S^{+}
\Gamma(U_1)^{\dagger}
\Gamma(U_0)^{\dagger}
\ket{E}
\\&\qquad\bra{\widetilde{E}}
\Gamma(U_0)\Gamma(\noise)
\Gamma(U_1)\Gamma(\noise)
\ket{\widetilde{\varrho}}.
\end{align}
By using \(g+1\) gates,
we have that the filtering process leads to
\begin{equation}
p_{\mu, \noise}^{g} c_\mu^{\varrho, \widetilde{\varrho}}.
\end{equation}
Therefore,
by randomly sampling a sequence of gates and increasing the circuit depth,
we end up with a scheme to estimate the parameters of the fidelity
for the noise \(\noise\).
\section{SU(\(m\)) characters}\label{app:characters}

Proposition~7.4 of Ref.~\cite{Buergisser2000a} gives an explicit formula for the
character in terms of a determinant built from elementary symmetric polynomials.
Let~\(U\in \mathrm{SU}(m)\) with eigenvalues \(x_1,\ldots,x_m\),
so that~\(U=V\operatorname{diag}(x_1,\ldots,x_m)V^{\dagger}\).
Then the character is the Schur polynomial \(\mathcal{S}_\mu\)
(defined below)
evaluated at these eigenvalues:
\begin{equation}\label{eq:char-schur}
  \chi_\mu(U)=\mathcal{S}_\mu(x_1, \ldots, x_m).
\end{equation}
Let \(\mu'=(\mu'_1,\ldots,\mu'_{\mu_1})\) be the conjugate partition (transpose
Young diagram) of \(\mu\), where \(\mu'_j\) equals the number of boxes in the
\(j\)-th column of the Young diagram of \(\mu\) (equivalently, \(\mu'\) is obtained
by transposing the diagram of \(\mu\)).
This is not, in general, the tableau of the dual irrep \(\mu^{*}\) defined in
Appendix~\ref{app:decomposition}; the two coincide only in special cases where
\(\mu\simeq\mu^{*}\).
Giambelli’s formula~\cite{Fulton1996}
expresses \(\mathcal{S}_\mu\) as
\begin{equation}\label{eq:giambelli}
  \mathcal{S}_\mu=\det\!\left[\sigma_{\mu'_i+j-i}\right]_{1\le i,j\le \mu_1},
\end{equation}
where \(\sigma_k\) are the elementary symmetric polynomials,
\begin{multline}\label{eq:elementary-sym}
  \sigma_k(x_1,\ldots,x_m)\coloneqq
  \sum_{1\le i_1<\cdots<i_k\le m} x_{i_1}\cdots x_{i_k},
  \\
  \qquad \sigma_0=1,\ \sigma_k=0\ \text{for }k>m.
\end{multline}
For example, \(\sigma_1=\sum_i x_i\) and \(\sigma_2=\sum_{i<j} x_i x_j\), while
\(\mathcal{S}_{(1)}=\sigma_1\), \(\mathcal{S}_{(1,1)}=\sigma_2\), and \(\mathcal{S}_{(2)}=\sigma_1^2-\sigma_2\).
For instance, for \(m=2\) and \(U\in \mathrm{SU}(2)\) with eigenvalues \(x_1,x_2\),
the character of the fundamental irrep \(\mu=(1,0)\) is
\(\chi_{(1,0)}(U)=\mathcal{S}_{(1)}=x_1+x_2\); the antisymmetric irrep \(\mu=(1,1)\)
has \(\chi_{(1,1)}(U)=\mathcal{S}_{(1,1)}=x_1x_2\); and the symmetric irrep \(\mu=(2,0)\)
has 
\begin{equation}\label{eq:useful-char}
\chi_{(2,0)}(U)=\mathcal{S}_{(2)}=x_1^2+x_2^2+x_1x_2.
\end{equation}

\end{document}